**Data-Driven Approach for Noise Reduction in Pressure-Sensitive Paint Data Based on Modal Expansion and Time-Series Data at Optimally Placed Points**

Tomoki Inoue[1, ‡], Yu Matsuda[1, 2, ‡ *], Tsubasa Ikami[3, 4], Taku Nonomura[3], Yasuhiro Egami[5], Hiroki Nagai[4]

1. Department of Modern Mechanical Engineering, Waseda University, 3-4-1 Ookubo, Shinjuku-ku, Tokyo, 169-8555, Japan

2. Japan Science and Technology Agency, PRESTO, Saitama 332-0012, Japan

3. Department of Aerospace Engineering, Tohoku University, 2-1-1 Katahira, Aoba-ku, Sendai, Miyagi-prefecture 980-8577, Japan

4. Institute of Fluid Science, Tohoku University, 2-1-1 Katahira, Aoba-ku, Sendai, Miyagi-prefecture 980-8577, Japan

5. Department of Mechanical Engineering, Aichi Institute of Technology, 1247 Yachigusa, Yakusa-Cho, Toyota, Aichi-prefecture 470-0392, Japan

‡ These authors contributed equally.

* Corresponding Author: y.matsuda@waseda.jp

**Abstract**

We propose a noise reduction method for unsteady pressure-sensitive paint (PSP) data based on modal expansion, the coefficients of which are determined from time-series data at optimally placed points. In this study, the proper orthogonal decomposition (POD) mode calculated from the time-series PSP data is used as a modal basis. Based on the POD modes, the points that effectively represent the features of the pressure distribution are optimally placed by the sensor optimization technique. Then, the time-dependent coefficient vector of the POD modes is determined by minimizing the difference between the time-series pressure data and the reconstructed pressure at the optimal points. Here, the coefficient vector is assumed to be a sparse vector. The advantage of the proposed method is a self-contained method, while existing methods use other data such as pressure tap data for the reduction of the noise. As a demonstration, we applied the proposed method to the PSP data measuring the Kármán vortex street behind a square cylinder. The reconstructed pressure data is agreed very well with the pressures independently measured by pressure transducers.

# 1. Introduction

The measurement of pressure distribution on a model of interest is important for a deep understanding of the flow field and for designing effective aerodynamic shapes. As a pressure distribution measurement technique, pressure-sensitive paint (PSP) has received much attention. [1-4] The pressure measurement using PSP is based on the oxygen quenching of the phosphorescence emitted from the dye included in the PSP layer. [2, 3, 5] Since the pressure distribution on a model can be obtained by measuring the emission intensity distribution by a camera, PSP has been applied to a variety of aerodynamic testing such as flows around wind tunnel models, [3, 6-12] around rotating objects, [13-17] in low-pressure range, [18-20] and inside a micro-duct. [21-24] As described in these literatures, demand for unsteady pressure measurement increases. Many studies have been conducted to improve the time response of PSP, called fast-responding PSP or fast-PSP. [7, 25-30] The application of fast-PSP to unsteady flow with small pressure fluctuation (less than 0.1 kPa) remains a frontier of PSP technique, because it is difficult to detect the small intensity variation due to the small pressure fluctuation with the short camera exposures. In other words, unsteady PSP measurement technique suffers from a poor signal-to-noise ratio (SNR). The phase-locked measurement [31, 32] and the Heterodyne method [33] were proposed to achieve a high SNR measurement. It should be noted that these methods are only applicable to periodic phenomena. In these days, data-driven post-processing methods have been intensively studied. The noise reduction method based on singular value decomposition (SVD), which is also known as proper orthogonal decomposition (POD), is applied to the data-processing for PSP images [34, 35]. In this method, noise reduction was conducted by truncating the SVD (POD). The dynamic mode decomposition (DMD),[36] Kalman-filter DMD,[37] and extended-Kalman-filter-based DMD [38] have also been applied. There are many choices for the truncation rank of the POD, which also relates to the DMD calculation, and for the mode selection of the DMD to reconstruct the pressure distribution data. As a noise reduction method based on the combination of PSP and the data obtained by other methods, the PSP data was phase-averaged based on the phase detected by pressure tap.[39] This method is a very powerful method for periodic phenomena. The application of the POD to the PSP images was also proposed.[40, 41] The coefficients for each POD mode to reconstruct the pressure distribution were determined from the comparison with the microphone data.

The exploring optimal sensor placement [42-44] for signal reconstruction in a tailored library has gathered attention in control theory. The required number of sensors can be dramatically reduced for the reconstruction using optimally placed sensors and the library. Since there are $n!/(n-q)!q!$ possible choices of $q$ sensor points for $n$-dimensional state, several approximation algorithms have been proposed [43, 45-47]. The sparse processing method based on the optimal sensor placement is applied to particle image velocimetry (PIV) for the real time flow observation. [48]

We proposed a noise reduction method based on modal expansion, the coefficients of which are determined from the time-series data at optimally placed points (sensors). We choose the POD as a modal basis. The optimal placement of points is estimated from the POD modes calculated from the obtained PSP images. The optimal POD modes and their coefficients are selected to fit the pressure data at optimally placed points by means of sparse modeling, while the distribution is reconstructed by the least-squares method in the existing method.[43, 44, 48] Based on the selected POD modes and the coefficients, the noise reduced pressure distribution can be reconstructed. We apply the proposed method to a wake flow behind a square cylinder and compare the result with the pressure data independently measured by pressure transducers. This method is a self-contained method that does not require other data such as the pressure tap and the microphone data used in existing methods [35, 39-41].

**2. Pressure-Sensitive Paint**

The pressure distribution measurement technique using PSP is based on the oxygen quenching of phosphorescence. PSP coating is composed of dye molecules and a binder material used to fix the dye molecules to a model surface of interest. When PSP coating is illuminated with UV/blue light, the dye molecules are excited and emit phosphorescence. The luminescent intensity decreases with an increase in partial pressure of oxygen or air pressure due to the oxygen quenching. By detecting the luminescent intensity distribution with a camera, the pressure distribution on the surface is obtained. The pressure is deduced from the Stern-Volmer relation [2, 3],

$$\frac{I_{\text{ref}}}{I} = A + B\frac{p}{p_{\text{ref}}} \qquad (1)$$

where $I$ and $p$ are the luminescent intensity and pressure, respectively. The subscript, ref, indicates a reference condition. An atmospheric condition is used as a reference condition in this study. Then, the $I_{\text{ref}}$ is the luminescent intensity of the PSP at an atmospheric condition $p_{\text{ref}}$. The constants $A$ and $B$ are the Stern-Volmer constants satisfying the constraint condition of $A + B = 1$ [49, 50].

In this study, we used the fast-PSP coating proposed in the previous study.[30] Tris(4,7-diphenyl-1,10-phenanthroline) ruthenium(II) dichloride (Ru(dpp)$_3$) was used as the dye of our PSP coating. The fast-PSP was prepared by mixing hydrophilic titanium(IV) oxide (TiO$_2$) particle (diameter of particle: 15 nm) with room-temperature-vulcanizing silicone at the particle mass content of 80%. The Stern-Volmer coefficients of our PSP coating were $A = 0.27$ and $B = 0.73$, respectively.

More detailed characteristics of the PSP can be found elsewhere. [30]

## 3. Proposed Method

The time-series pressure distribution $p(s, k_h \Delta t)$ is deduced from the intensity distribution of PSP coating. The pressure distribution at the position vector $s$ is measured at $k_h \Delta t$, where $\Delta t$ is the time interval of the measurement, $k_h$ denoted as the subscript represents $m$ consecutive integers such as $k_1, k_2, \ldots, k_m$, and $m$ is the total number of the measurements. The pressure distribution $p(s, k_h \Delta t)$ is reshaped into a column vector as $p_{k_h}$. Then, the size of vector $p_{k_h}$ is $n = n_h \times n_v$, where $n_h$ and $n_v$ are the number of horizontal and vertical pixels of image, respectively. The observation data matrix $X$ with $m$ snapshots is defined as follows,

$$X = [p_{k_1}, p_{k_2}, \cdots, p_{k_m}]. \tag{2}$$

The SVD of the observation data matrix $X \in \mathbb{R}^{n \times m}$ provides the following matrix decomposition:

$$X = U \Sigma V^\mathsf{T}, \tag{3}$$

where $U \in \mathbb{R}^{n \times n}$ and $V \in \mathbb{R}^{m \times m}$ are unitary matrices and are called left and right singular matrix, respectively. The superscript $\mathsf{T}$ denotes the transpose. The matrix $\Sigma \in \mathbb{R}^{n \times m}$ is a diagonal matrix with singular values. The matrix $U$ represents the POD modes as

$$\boldsymbol{U} = [\boldsymbol{\psi}_1, \boldsymbol{\psi}_2, \cdots, \boldsymbol{\psi}_m]. \tag{4}$$

where $\boldsymbol{\psi}_j$ indicates $j$-th POD mode. The data can be approximately represented by a rank-$r$ truncated SVD ($r < n$) as

$$\boldsymbol{X} \approx \widetilde{\boldsymbol{U}}\widetilde{\boldsymbol{\Sigma}}\widetilde{\boldsymbol{V}}^\top, \tag{5}$$

where $\widetilde{\boldsymbol{U}} \in \mathbb{R}^{n \times r}$, $\widetilde{\boldsymbol{\Sigma}} \in \mathbb{R}^{r \times r}$ and $\widetilde{\boldsymbol{V}} \in \mathbb{R}^{m \times r}$ are truncated matrices and $\widetilde{\boldsymbol{U}}$ is the truncated POD modes.

The optimized sensor placement problem is based on the idea that the limited number of the data at optimally placed points (the data of optimally places sensors) will approximately represent full state information when a high-dimensional state has a low-rank representation in a tailored basis such as the POD basis. This problem expressed as follows [43]:

$$\begin{aligned} \boldsymbol{y} &\approx \boldsymbol{C}\widetilde{\boldsymbol{U}}\boldsymbol{a} \\ &= \boldsymbol{\Theta}\boldsymbol{a} \end{aligned} \tag{6}$$

where $\boldsymbol{y} \in \mathbb{R}^q$, $\boldsymbol{C} \in \mathbb{R}^{q \times n}$, $\widetilde{\boldsymbol{U}} \in \mathbb{R}^{n \times r}$, and $\boldsymbol{a} \in \mathbb{R}^r$ are the observation vector, a sparse sensor matrix, the POD modes, and the coefficient vector for the POD modes, respectively. Here, $q$ and $r$ are the number of points (sensors) and the POD mode, respectively. Each row vector of $\boldsymbol{C}$ has a single unity element indicating the sensor position. The matrix $\boldsymbol{\Theta}$ is defined as $\boldsymbol{\Theta} \coloneqq \boldsymbol{C}\widetilde{\boldsymbol{U}}$. The problem is to find a combination of $\boldsymbol{C}$, $\boldsymbol{y}$, and $\boldsymbol{a}$ that satisfies Eq. (6). In this study, we use the determinant-based greedy algorithm proposed by Refs. [45, 46] to estimate $\boldsymbol{C}$. Eq. (6) is

approximately solved as

$$\tilde{a} = \Theta^+ y = \begin{cases} \Theta^\top(\Theta\Theta^\top)^{-1}y, & (q < r) \\ (\Theta^\top\Theta)^{-1}\Theta^\top y, & (q \geq r) \end{cases} \quad (7)$$

where $\Theta^+$ is the pseudo inverse of $\Theta$. The error between $a$ and $\tilde{a}$ is minimized by maximizing the determinant of $\Theta\Theta^\top$ for $q < r$ or $\Theta^\top\Theta$ for $q \geq r$ [45, 46]. Then, the sensor positions $C$ is deduced from the POD modes $\tilde{U}$ and $\Theta$ maximizing the determinant. For more details, the reader is referred to Refs. [45, 46]

Next, the time-series pressure data at optimally placed points, which are estimated by the optimized sensor placement problem, is considered. This time-series data is extracted from the pressure distribution $p_{l_h}$ at time $\tau_h = l_h \Delta t$, where $l_h$ is consecutive integers as $l_h = l_1, l_2, \cdots, l_N$. It is noted that the pressure data $p_{l_h}$ can be obtained at a different time than the data $p_{k_h}$ used in the calculation of the POD modes and the optimal sensor placement; that is, $p_{l_h}$ is not contained in $p_{k_h}$. Since each pressure data contains a large amount of noise due to unsteady PSP measurement with a short camera exposure and small pressure fluctuation, denoising process is required. Though there are many sophisticated denoising methods such as the Kalman filter family and frequency-domain filtering, a spatial mean filter around the optimal point is used as a simple filter in this study. The filtered time-series pressure data at the optimally placed points is

denoted as

$$\boldsymbol{\Phi} = [\boldsymbol{\phi}_{l_1}, \boldsymbol{\phi}_{l_2}, \cdots, \boldsymbol{\phi}_{l_N}]. \tag{8}$$

where the vector $\boldsymbol{\phi} \in \mathbb{R}^q$ is generated by arranging the pressure data at the optimal points.

The denoised pressure distribution $\widetilde{\boldsymbol{X}} \in \mathbb{R}^{n \times l_N}$ can be reconstructed by the POD modes as follows

$$\widetilde{\boldsymbol{X}} \approx \widetilde{\boldsymbol{U}} \boldsymbol{A} \tag{9}$$

where $\boldsymbol{A} \in \mathbb{R}^{n \times r}$ is the amplitude matrix for the POD modes. Eq. (9) is also denoted as

$$\widetilde{\boldsymbol{p}}_{l_h} = \boldsymbol{\alpha}_{l_h} \widetilde{\boldsymbol{U}} = \sum_{i=1}^{r} \alpha_{l_h,i} \boldsymbol{\psi}_i \tag{10}$$

where $\boldsymbol{\alpha}_{l_h} = (\alpha_{l_h,1}, \alpha_{l_h,2}, \cdots, \alpha_{l_h,r})^T$ and the amplitude $\alpha_{l_h,i}$ for the $i$-th POD mode at the time $l_h \Delta t$ is unknown. In this study, the amplitude $\boldsymbol{\alpha}_{l_h}$ is determined by minimizing the difference between the pressure data $\boldsymbol{\phi}_{l_h}$ and the reconstructed pressure $\boldsymbol{C} \widetilde{\boldsymbol{p}}_{l_h}$ at the optimally placed points. Assuming that the amplitude $\boldsymbol{\alpha}_{l_h}$ is sparse, we can determine each amplitude $\boldsymbol{\alpha}_{l_h}$ by the LASSO (least absolute shrinkage and selection operator) [51-53] as follows:

$$\boldsymbol{\alpha}_{l_h} = \underset{\boldsymbol{\alpha}_{l_h}}{\operatorname{argmin}} \frac{1}{2q} \left\| \boldsymbol{\phi}_{l_h} - \boldsymbol{C} \boldsymbol{\alpha}_{l_h} \widetilde{\boldsymbol{U}} \right\|_2^2 + \lambda \left\| \boldsymbol{\alpha}_{l_h} \right\|_1 \tag{11}$$

where $\|\blacksquare\|_2$ and $\|\blacksquare\|_1$ indicate the $\ell_2$ and the $\ell_1$ norm of $\blacksquare$, respectively. The parameter

$\lambda$ weights the importance of sparsity. The ADMM (alternating determination method of multipliers) [53, 54] algorithm is used to solve Eq. (11). The outline of the proposed method is shown in Fig. 1.

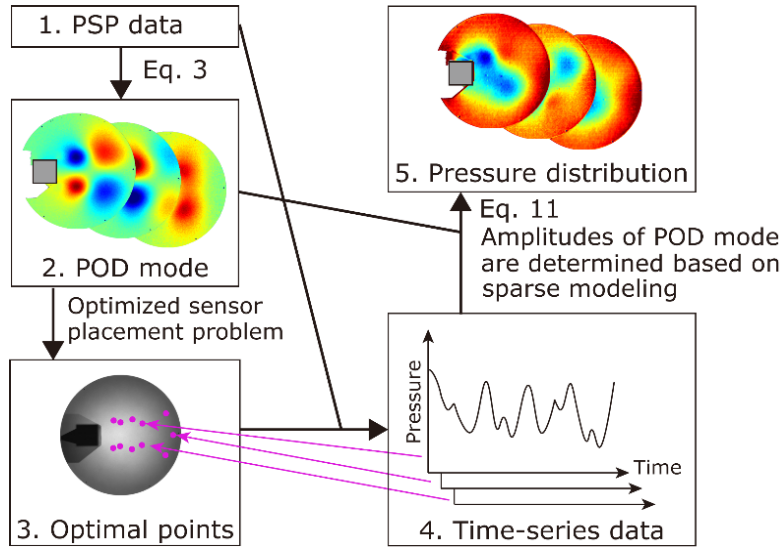

Figure 1  Outline of proposed method

**4. Experimental method**

The proposed method is applied to the PSP data to validate the denoising performance. In this study, we conducted unsteady PSP measurement at the small low-turbulence wind tunnel at the Institute of Fluid Science, Tohoku University. A brief explanation is provided below. More details of the experiment are described elsewhere.[30] The schematic illustration of the experimental setup is shown in Fig. 2. The wind tunnel has a test section with 0.29 m width and 0.9 m length. A square cylinder (width: 40 mm, length: 40 mm, and height: 100 mm) was placed 170 mm from the elliptic leading edge of a flat plate (width: 355 mm and length: 600 mm) installed at the test section. As shown in Fig. 2b, the PSP coating described in Sec. 2 was applied to a turntable with a diameter of 250 mm. Two LED devices with a central wavelength of 462 nm (IL-106B, HARDsoft microprocessor systems, Poland) were used as illumination light sources of the PSP. A cold filter (SC0451, Asahi Spectra, Japan) and a heat absorption filter (HAF-50 S-30 H, SIGMAKOKI, Japan) were placed in front of the LED lens to remove infrared light from the LED device. The emission light from the PSP was detected by a high-speed camera (FASTCAM Nova,

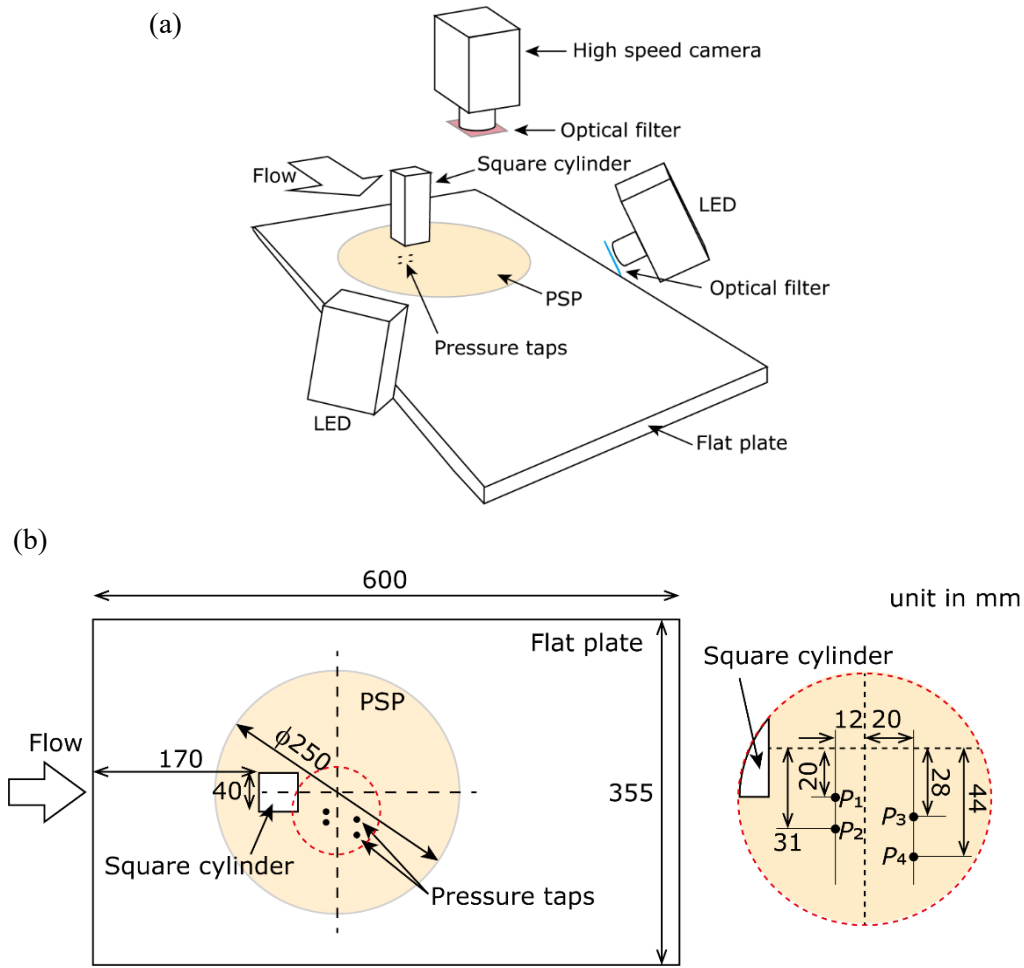

Figure 2 Experimental setup. (a) schematic illustration of PSP measurement, (b) position of pressure taps ($P_1$, $P_2$, $P_3$, and $P_4$)

Photron, Japan) with a lends (Nikkor 50mm f/1.4, Nikon, Japan) through a band-pass filter of 640 ± 50 nm (PB0640, Asahi Spectra). The exposure time was set to 1/3000 s. The frame rate of the camera was 1000 fps, and the time interval of the images is $\Delta t = 1/1000$ s. In this study, we measured the pressure distribution at a mean velocity of 50 m/s; thus, the "wind-on" images were captured at a mean velocity of 50 m/s and "wind-off" images were immediately captured after the wind tunnel was stopped.

The pressures ($P_1$, $P_2$, $P_3$, and $P_4$) downstream of the square cylinder were measured by pressure transducers (XCQ-062-50a, Kulite, USA) at the pressure taps (diameter: 0.5 mm) as shown in Fig. 2b. The pressures were recorded with 10 kHz frequency. The obtained data was moving averaged and downsampled to synchronize the 1 kHz camera acquisition.

## 5. Results and discussion

The time-series pressure distribution $\boldsymbol{p}_{k_h}$ ($k_1, k_2, \ldots, k_{8192}$) and $\boldsymbol{p}_{l_h}$ ($l_h = l_1, l_2, \cdots, l_{512}$) were deduced from the wind-on and wind-off images by Eq. (1). It is noted that $\boldsymbol{p}_{k_h}$ and $\boldsymbol{p}_{l_h}$ were measured at different times as mentioned in Sec. 3. In this study, the pressure was normalized by the pressure at an atmospheric (reference and wind-off) pressure $p_{\text{ref}}$. We trimmed away the area

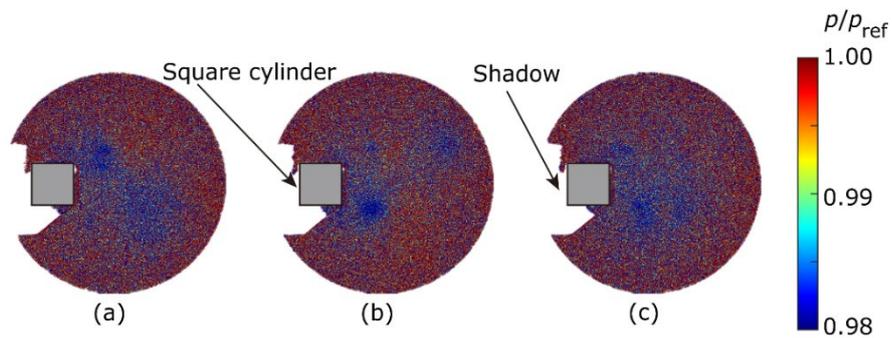

Figure 3  Typical measured pressure distribution before denoising. (a) pressure distribution at $l_{50}$, (b) pressure distribution at $l_{150}$, and (c) pressure distribution at $l_{250}$

out of the PSP from the obtained image. The size of the trimmed image was $780 \times 780$ pixels.

The POD modes and optimally placed points calculated from $\boldsymbol{p}_{k_h}$ are used to reduce the noise

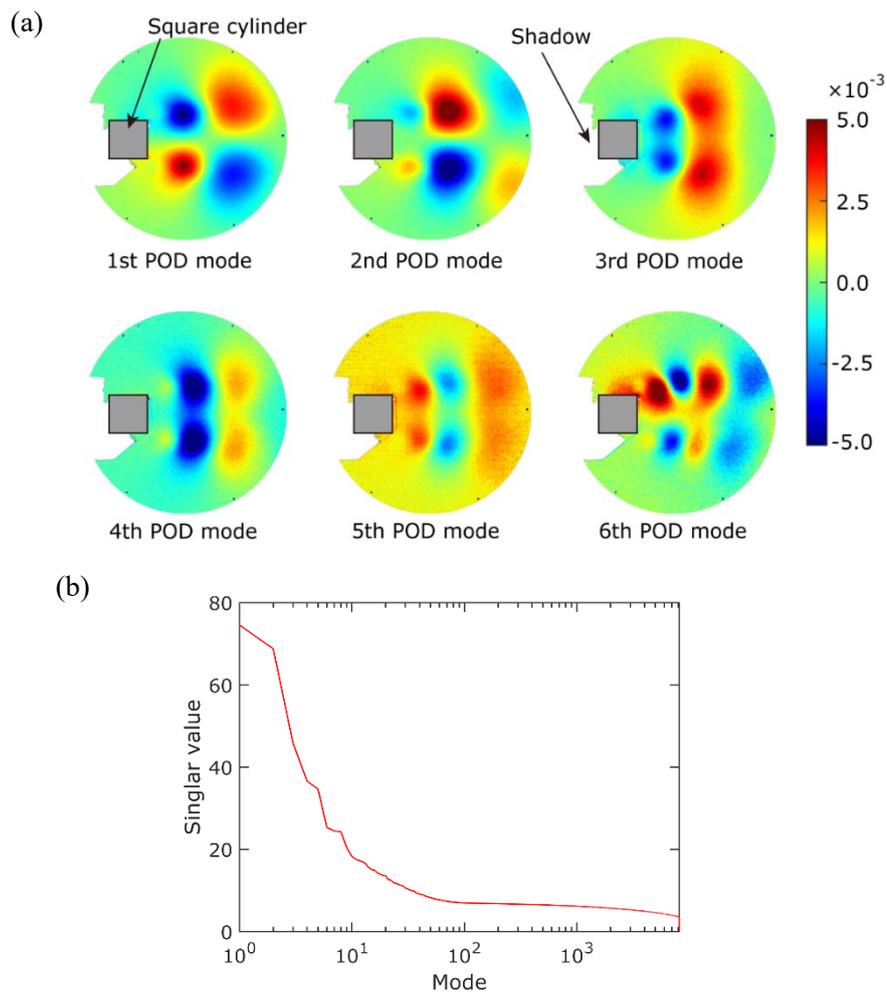

Figure 4  Calculated POD modes and singular values. (a)first 6 POD modes calculated from 8192 snapshots, (b) relation between singular value and mode

of $p_{l_h}$. Typical pressure distributions of $p_{l_h}$ are shown in Fig. 3. As shown in the figure, flow structures are hardly recognized due to the low signal-to-noise ratio (SNR) of the image. The pressure distribution data of 8192 snapshots is arranged to the matrix $X$ in the manner of Eq. (2). The POD modes $U$ were calculated from $X$, which is obtained by concatenating the snapshot pressure distribution $p_{k_h}$ as shown in Eq. (2). The first 6 POD modes are shown in Fig. 4 as examples. The small black circles are the markers for image alignment. The distributions are symmetrical/antisymmetrical with respect to the centerline along the flow direction for the first 5 modes. On the other hand, the distribution of the sixth POD mode is asymmetric. Such asymmetrical distributions are recognized at higher modes due to disturbances in amplitude and phase. These asymmetrical modes are not observed in ensemble averaging methods [39]. Based on the calculated POD modes, the optimally placed points were estimated. Though the truncation rank was calculated as 100 following the method proposed by Gavish and Donoho [55] in our condition, the rank $r = 128$ truncated POD modes $\tilde{U}$ was used for redundancy to calculate

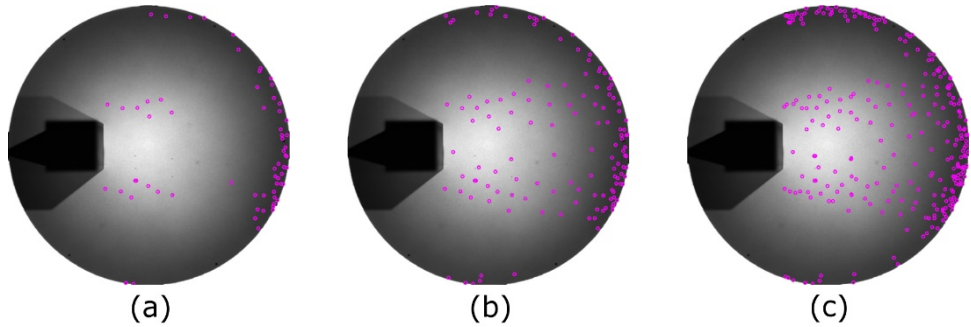

Figure 5 Estimated optimally placed points indicated by magenta circles
(a) number of the points $q = 64$, (b) $q = 128$, and (c) $q = 256$

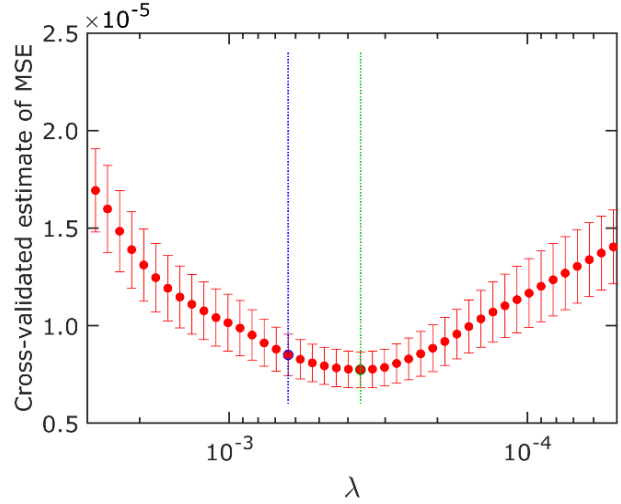
Figure 6 Typical result of cross-validated estimate of mean-squared error. The green and blue dotted lines indicate the minimum error and the one-standard-error location, respectively.

optimally placed points. In this study, we searched for the optimal points at the region downstream of the square cylinder. The number of optimal points $q$ were set to $q = 32, 64, 128, 256,$ and $512$. The estimated optimal points for $q = 64, 128,$ and $256$ are shown in Fig. 5. The places with large variations between each POD mode are selected.

Next, we prepared the time-series pressure data at the optimally placed points $\boldsymbol{\Phi}$, defied by Eq. (8), from $\boldsymbol{p}_{l_h}$. The spatial averaging filter of $11 \times 11$ pixels is applied, and the noise from the time-series pressure data is reduced. The large spatial averaging filter reduces the noise but also reduces the spatial resolution; thus, this relatively large spatial averaging filter is only applied to time-series pressure data. The amplitude vector $\boldsymbol{\alpha}_{l_h}$ are calculated from Eq. (11). We estimated the optimal choice of the regularization parameter $\lambda$ based on 10-fold cross-validation using the one-standard-error rule [52, 53, 56]. Figure 6 shows the typical result of the cross-validated estimate of mean square error (MSE). Then, the pressure distributions were reconstructed from Eq. (10). Typical pressure distributions reconstructed based on the number of optimally placed points $q = 128(= r)$ at the same time as Fig. 3 are shown in Fig. 7. The movies of the pressure distribution are shown in the Supplementary. For comparison, the pressure distributions reconstructed by the amplitudes estimated by the least-squares method ($\lambda = 0$ in Eq. (11)) are also shown. As shown in Fig. 7, the noise was significantly reduced in the proposed method compared with the least-squares method and the raw data (Fig. 3). The flow structure is clearly recognized in the proposed method. The pressures at the pressure taps are also compared with those measured by the pressure transducers as shown in Fig. 8. The pressures and the phase processed by the proposed method show good agreement with those measured by the pressure transducers. The root mean square error (RMSE) between each processed PSP data and the data measured by pressure transducers are shown in Fig. 9. For comparison, the PSP data with the spatial averaging filter of $11 \times 11$ pixels is shown. It is found that the RMSE of the proposed method are the smallest. Though the RMSE of the proposed method decreases with increasing the number of optimally placed points for the number of the points less than the truncation rank $r$ ($q < r$), the RMSE is almost constant for $q > r$. Thus, the number of the optimal points of $q = r$, $q = 128$ in this study, is enough to reconstruct the pressure distributions.

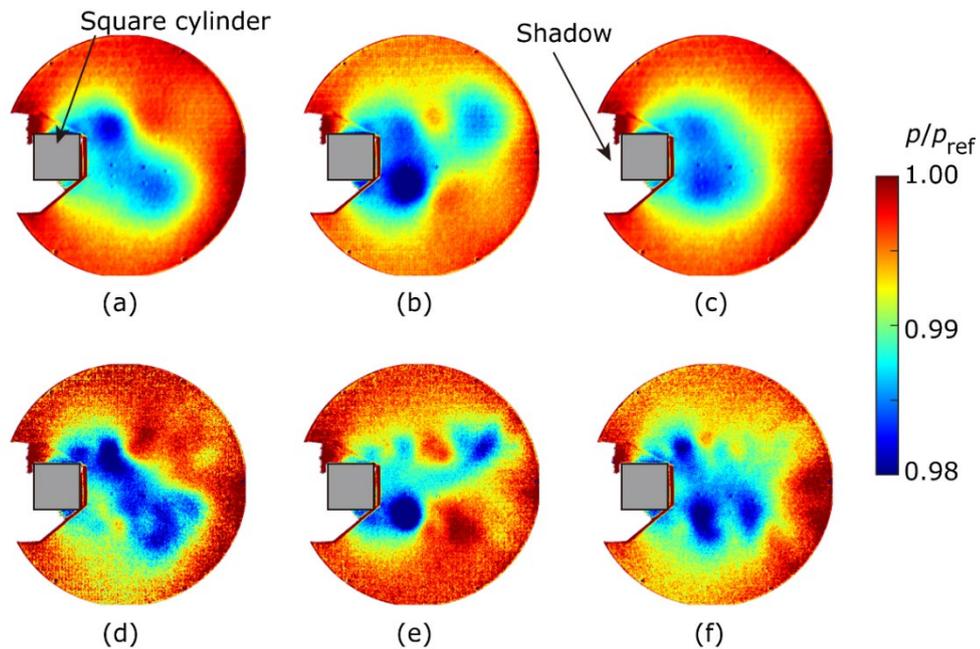

Figure 7 Pressure distributions after denoising based on the number of optimally placed points of $q = 128$. Typical pressure distribution denoised by the proposed method at (a) $l_{50}$, (b) $l_{150}$, and (c) $l_{250}$. Typical pressure distribution denoised by the least-squares method ($\lambda = 0$) at (d) $l_{50}$, (e) $l_{150}$, and (f) $l_{250}$.

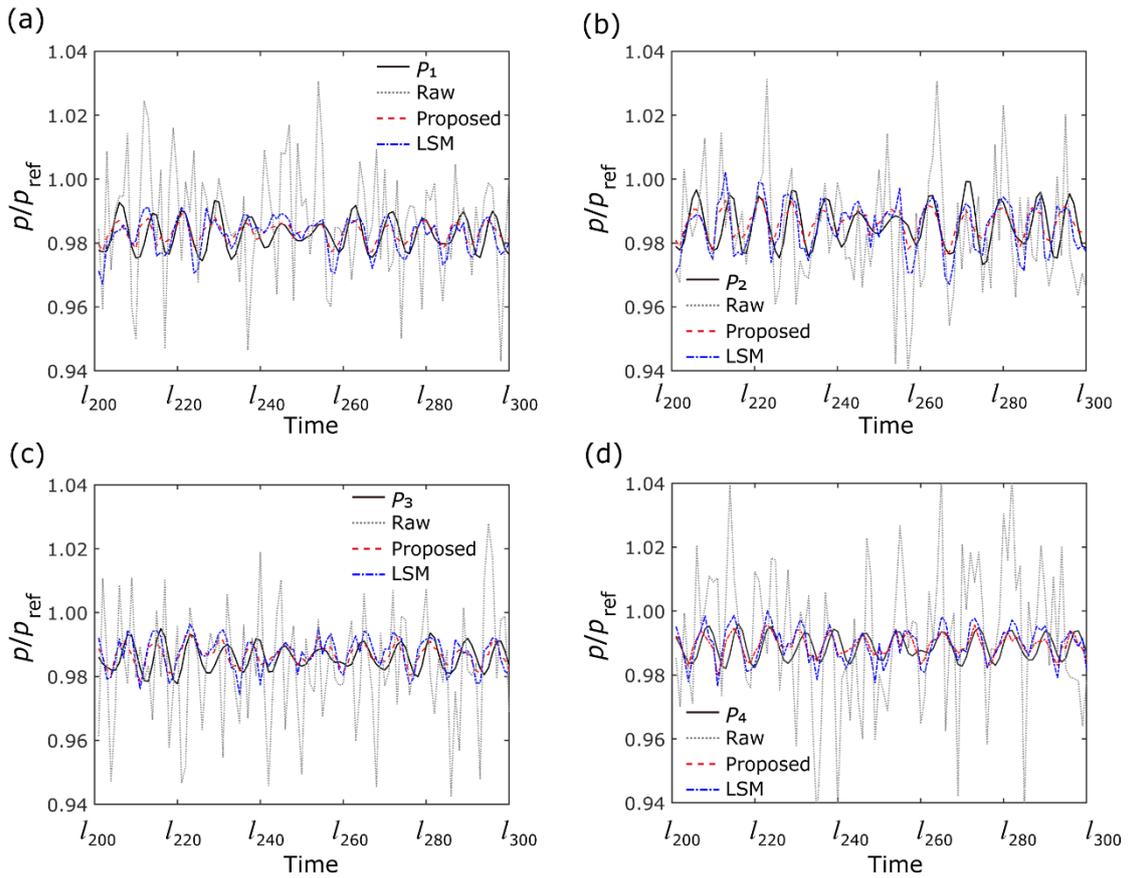

Figure 8  Comparison between pressures measured by PSP and those measured by pressure transducers through pressure taps. $P_1$, $P_2$, $P_3$, and $P_4$ indicate the pressures at the pressure taps (see also Fig. 2). Raw indicates PSP data without denoising. Proposed and LSM indicate the PSP data denoised by the proposed method and the least-squares method. The number of optimally placed points is $q = 128$.

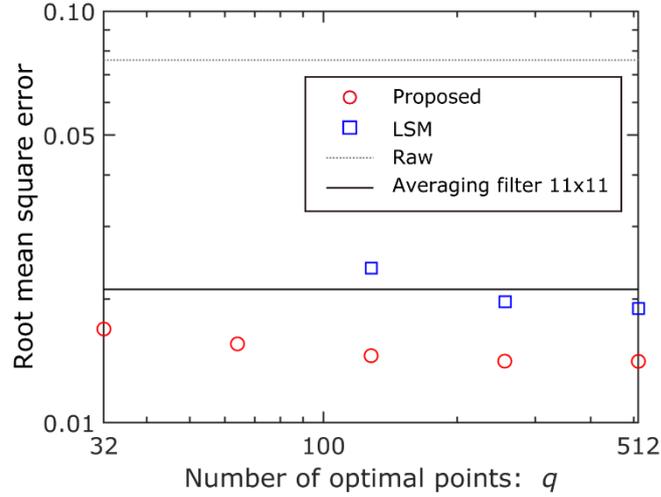

Figure 9 Root mean square error for the PSP data. The PSP data were denoised by the proposed method (Proposed), least-squares method (LSM), and the spatial averaging filter of $11 \times 11$ pixels. The raw data without denoising is also shown.

The estimated amplitude $\boldsymbol{\alpha}_{l_h}$ for $r = q = 128$ and $l_h = 150$ is shown in Fig. 10 as a typical example. This indicates that the pressure distribution can be reconstructed by a small number of the POD modes and some of the lower modes are not necessary for the reconstruction. The POD modes showing the model vibration [34] and the camera pattern [30] are not chosen in the manner of sparse modeling. Importantly, the pressure distribution with disturbances in amplitude and phase can be resolved by the proposed method, while such signals are indistinguishable by the previous method as a result of ensemble averaging.[39]

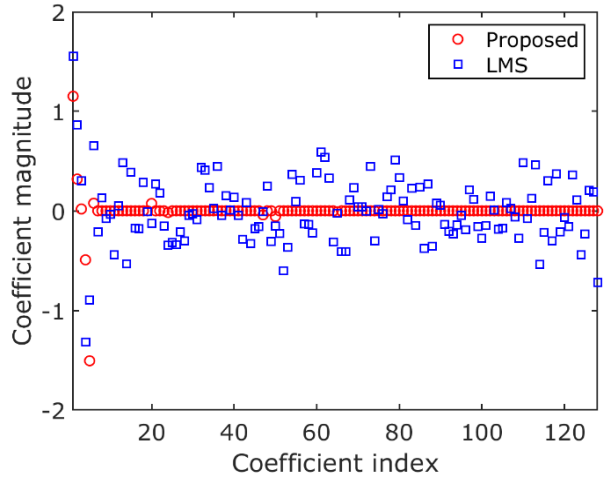

Figure 10 Estimated amplitude $\alpha_{l_h}$ for $r = q = 128$ and $l_h = 150$ by the proposed method (Proposed) and least-squares method (LSM).

## 6. Conclusion

We propose a noise reduction method for unsteady pressure-sensitive paint (PSP) data to resolve small pressure fluctuation. The proposed method is based on modal expansion whose coefficients are determined from optimally placed points data. In this study, the proper orthogonal decomposition (POD) mode is used as a modal basis. The optimally placed points, which effectively represent the features of the pressure distribution, are calculated based on the POD modes. The denoised data can be represented by the POD modes and the appropriate coefficients for each POD mode. The coefficients are determined by minimizing the difference between the filtered pressure data and the reconstructed at the optimally placed points. Assuming that the coefficients are sparse, we calculate the coefficients by the least absolute shrinkage and selection operator (LASSSO). As a demonstration, we applied the proposed method to experimentally measured pressure distribution induced by the Kármán vortex street behind a square cylinder. The pressure distribution was successfully reconstructed by the POD modes and the time-dependent coefficient vector. The reconstructed pressure data agree very well with the pressures independently measured by pressure transducers connected to pressure taps. The advantage of the proposed method is a self-contained method, while existing methods use other data such as pressure tap data to reduce the noise. Our proposed method is a promising method for noise reduction of PSP data.

**Acknowledgments**

The authors wish to thank Dr Yasuhumi Konishi, Mr Hiroyuki Okuizumi, Mr Hiroya Ogura, Mr Yuya Yamazaki, Mr Kakeru Kurane, during the wind tunnel testing at the Institute of Fluid Science, Tohoku University. We also gratefully appreciate Tayca corp. for providing titanium dioxide. This work was partially supported by JST, PRESTO Grant Number JPMJPR187A, Japan. Part of the work was carried out under the Collaborative Research Project of the Institute of Fluid Science, Tohoku University.

**Data Availability**

The data that support the findings of this study are available from the corresponding author upon reasonable request.